# Cloaking a nanolaser

Sergey Lepeshov[1], Andrey Vyshnevyy[2], and Alex Krasnok[3,*]

[1]*Department of Physics and Engineering, ITMO University, St. Petersburg, Russia*

[2]*Center for Photonics and 2D materials, Moscow Institute of Physics and Technology, Dolgoprudny, 141700, Russia*

[3]*Department of Electrical and Computer Engineering, Florida International University, Miami, FL 33174, USA*

**E-mail: akrasnok@fiu.edu*

**Abstract**

Light emitters are bound to strongly interact with light through enhanced absorption and scattering, which imposes limitations on the design and performance of photonic devices such as solar cells, nanoantennas, and (nano) lasers. Overcoming these limitations forces the use of ineffective nonreciprocity approaches or separation of radiation and scattering in the frequency or time domain. A design that combines the properties of an efficient emitter in one state and the property of being invisible in another state is vital for various applications. In this work, we propose a nanolaser design based on a semiconductor nanoparticle with gain coated by a phase transition material ($Sb_2S_3$), switchable between lasing and cloaking (nonscattering) states at the same operating frequency without change in pumping. The operation characteristics of the nanolaser are rigorously investigated. The designed nanolaser can operate with optical or electric pumping and possesses attributes of a thresholdless laser due to the high beta-factor and strong Purcell enhancement in the strongly confined Mie resonance mode. We design a reconfigurable metasurface composed of lasing-cloaking metaatoms that can switch from lasing to a nonscattering state in a reversible manner.

## Introduction

As the miniaturization of optical devices from large-scale to nanoscale photonic structures has become a current trend in advanced photonics, a new type of nanoscale coherent emitters called nanolasers has been proposed. In 2003, a pioneer work by David Bergman and Mark Stockman proposed the concept of surface plasmon amplification by stimulated emission of radiation (SPASER) which marked the beginning of the era of spasers and nanolasers[1]. Since that time, spasers and nanolasers have obtained a variety of forms and designs[2]. A noble metal nanoparticle (NP) covered by active media has been established as the most feasible design for spasers and nanolasers, providing a good overlap between the material gain and lasing mode[3–7]. Further, semiconductor-based nanolasers have appeared as low-threshold and efficient counterparts of plasmonic spasers[8,9]. It is noteworthy that perovskites and quantum dot (QD) inclusions have received the greatest recognition due to the high material gain and optical activity provided by these materials[10–13]. Nanolasers are successfully used for sensing, biological probing, super-resolution imaging techniques and vortex beam generation[4]. Furthermore, the development of nanolasers approaches on-chip integrated optical interconnects and all-optical data processing.

In existing nanolaser architectures, there is no possibility of tuning between different scattering states, being limited to switching from narrowband laser radiation to broadband resonant or nonresonant scattering, achieved by adjusting the pump intensity. One system of such kind, the so-called anapolar laser, has recently attracted a great deal of attention because the anapolar state is likely to suppress lasing threshold intensity in geometries with optical pumping[14,15]. However, in order to transform this nanolaser to another scattering mode, for instance, a mode with suppressed scattering provided by the anapole, it is necessary to turn off the optical pumping, that is, to significantly change the system. An interesting question then arises: can a nanolaser be cloaked without changing its pumping conditions?

Indeed, one of the most fundamental theorems, the reciprocity theorem, implies that an efficient light emitter simultaneously interacts strongly with its light through enhanced absorption and scattering[16–20]. For example, antennas, no matter whether microwave, THz or optical, are forced to transmit and receive with the same efficiency to/from the same direction[21–23]. The principle of detailed balance, the direct consequence of reciprocity, being applied to thermal emission, gives rise to Kirchhoff's law of equivalence of spectral emissivity and spectral absorptivity of any system in its thermal equilibrium[24]. A direct bandgap photodiode effectively converting light into electrical energy can emit light working as a light-emitting diode (LED) element and vice versa regardless of its constitutive materials. The laser, being an active and nonlinear system that does not fully obey the reciprocity theorem, is also known to interact strongly with incoming (e.g., back-reflected) light, which forces us to use optical isolators to prevent this reflection. As a consequence of this fact, a nanolaser is known to exhibit a strongly enhanced scattering in the quasi-linear regime at the lasing threshold[13].

The fact that light-emitting systems interact strongly with this radiation enables dual devices, e.g., receiving-transiting antennas and radar systems[21], emitting-absorbing nanoantennas[25], full-duplex telecommunications, optically pumped laser-LED[26], to name a few. However, this circumstance implies fundamental restrictions implied on the performance of optical devices over the entire spectral range. For example, solar cells that efficiently absorb solar energy are known to lose in conversion efficiency due to increased thermal radiation, and conversely, poorly emitting solar cells sacrifice solar energy absorption[27–29]. A perfect solar cell would also be an ideal LED, but reciprocity dictates that in reality, neither device exists[20]. The existing approaches to overcome these limitations of reciprocal systems rely on either a strong magnetic field bias created by lossy and bulky materials, the use of intrinsically weak optical nonlinearity, or a rarely applicable and often impractical method of time variation. Another approach is distinguishing the effects of light emission and absorption/scattering in frequency or time domain. An example of the latter is LIDAR, where a transceiver radiates a signal into some direction, and a receiver absorbs the reflected signal arriving from the same direction with a time delay proportional to the altitude or distance. In such systems, one has to ensure that the effects of light absorption/scattering and light emission are spaced sufficiently in frequency or time.

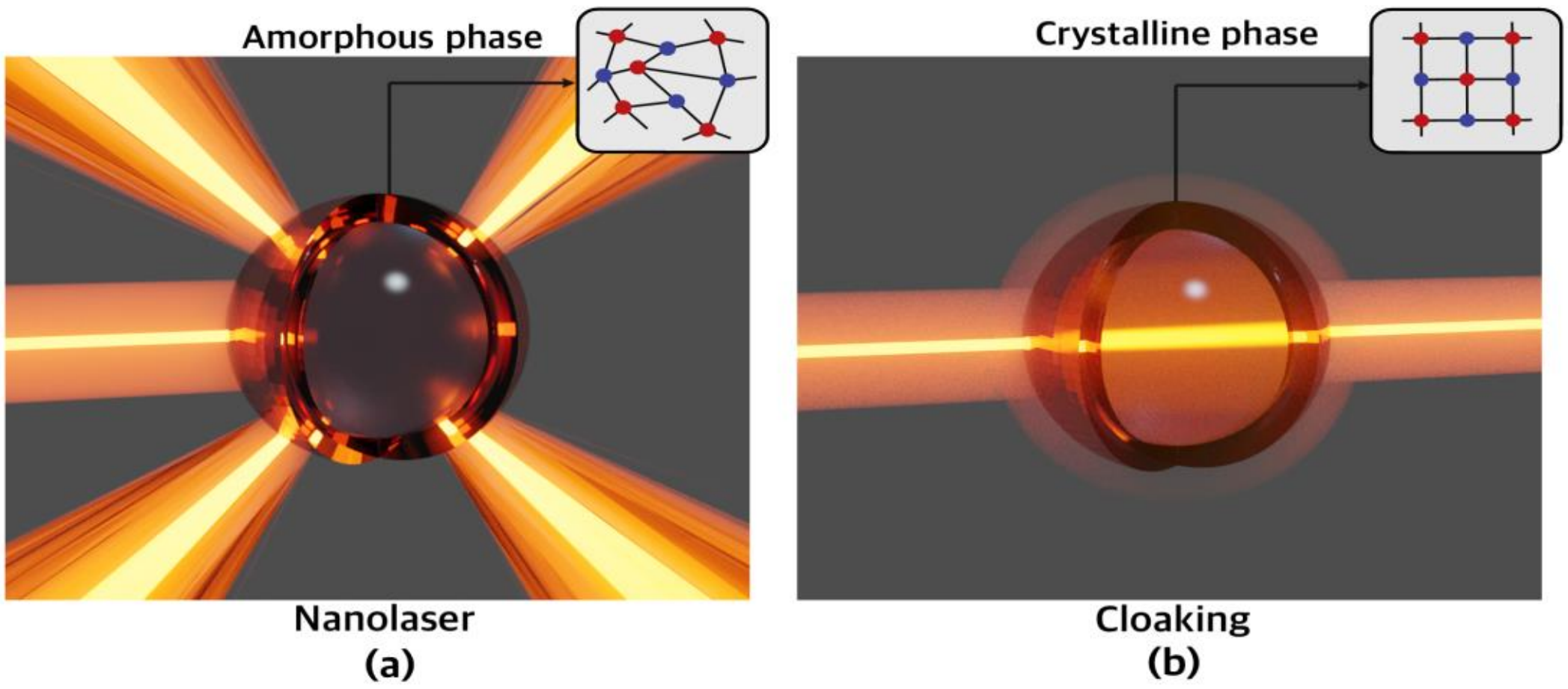

**Figure 1. Cloaking a nanolaser.** Illustration of the core-shell nanostructure supporting reversible switch between **(a)** nanolaser regime and **(b)** cloaking (anapole) regime at the same frequency.

In this work, we propose an approach that allows switching a nanolaser between emitting and cloaking (non-scattering) modes at the same operating frequency. The design is a core-shell nanostructure consisting of a semiconductor nanoparticle (NP) coated by a film of a phase change material (PCM). The PCM shell provides universal refractive index adjustment associated with the transition between amorphous and crystalline states. This nanostructure allows to reversibly switch between two optical states: coherent light emission [Fig. 1(a)] and cloaking [Fig. 1(b)]. The cloaking phase of the nanostructure is associated with the excitation of an anapole state[30,31]. We investigate the dynamics of the proposed tunable nanolaser in the frequency domain and input-output characteristics in the stationary state. Using this single lasing-cloaking metaatom, we design a reconfigurable metasurface with the ability to switch from laser radiation to a nonscattering state in a reversible manner. We demonstrate that such a wisely tailored nanolaser can indeed be cloaked and then brought back to the laser state without changing its pumping conditions.

## Results and discussion

We consider a spherical core-shell nanostructure made of a semiconductor core with material gain and an antimony trisulfide ($Sb_2S_3$) shell. $Sb_2S_3$ is a prospective phase-change material with a high dielectric constant, low loss, and strong tunability in the visible range[32,33]. The permittivity dispersion of $Sb_2S_3$ (see Supplementary Information) indicates the broadband transparency of this material above the wavelength of 600 nm in the amorphous phase. We assume that the permittivity of the semiconductor core obeys the Drude-Lorentz model[13]:

$$\varepsilon_c(\omega) = \varepsilon_\infty - \frac{f\omega_0^2}{\omega_0^2 - \omega^2 - i\Gamma\omega} \tag{1}$$

where $\varepsilon_\infty$ = 13.5 is the dispersionless part of the permittivity, $\omega$ is the angular frequency of light, $\omega_0/2\pi$ = 474 THz is the resonant frequency of dipole transitions, $\Gamma = 10^{13}$ $s^{-1}$ is the polarization decay rate, and $f$ is the oscillator strength related to the material gain $g$ and speed of light $c$ as[34] $f = gc\Gamma\sqrt{\varepsilon_\infty}/\omega_0^2$. Chosen parameters are achievable by QDs and perovskites at room temperature[13]. The permittivity of 13.5 is typical for high-index semiconductors like GaAs, GaP, InAs, and others. The resonant frequency $\omega_0$ is optimized to achieve maximum spectral overlap with the laser mode.

This model allows us to study the core-shell nanostructure in optically passive ($g \le 0$) and active ($g \ge 0$) regimes. To this end, we chose the core radius of $R_c$ = 109 nm and the radius of the shell of $R_s$ = 125 nm. The optical response of the spherically symmetric core-shell nanostructure can be described by the generalized Mie theory[35]. The scattering cross-section (SCS) spectrum of the nanostructure can be calculated from the Mie multipole expansion[36,37]: $\mathrm{SCS} = \frac{2}{(kR_s)^2}\sum_{l=1}^{\infty}(2l+1)(|a_l|^2 + |b_l|^2)$, where $l$ is the multipole order, $k = \omega/c$ is the vacuum wavenumber, $a_l$ and $b_l$ are the frequency-dependent electric and magnetic multipole scattering coefficients. According to the scattering theory, the optical properties of a system are defined by the eigenmodes of the system, which appear as poles of the scattering matrix coefficients at the corresponding complex eigenfrequencies[38]. Thus, to study the eigenmodes, eigenfrequencies, and their dynamics in the frequency domain, we calculate the scattering cross-section of the core-shell nanostructure in the complex frequency plane ($\omega = \omega' + i\omega''$).

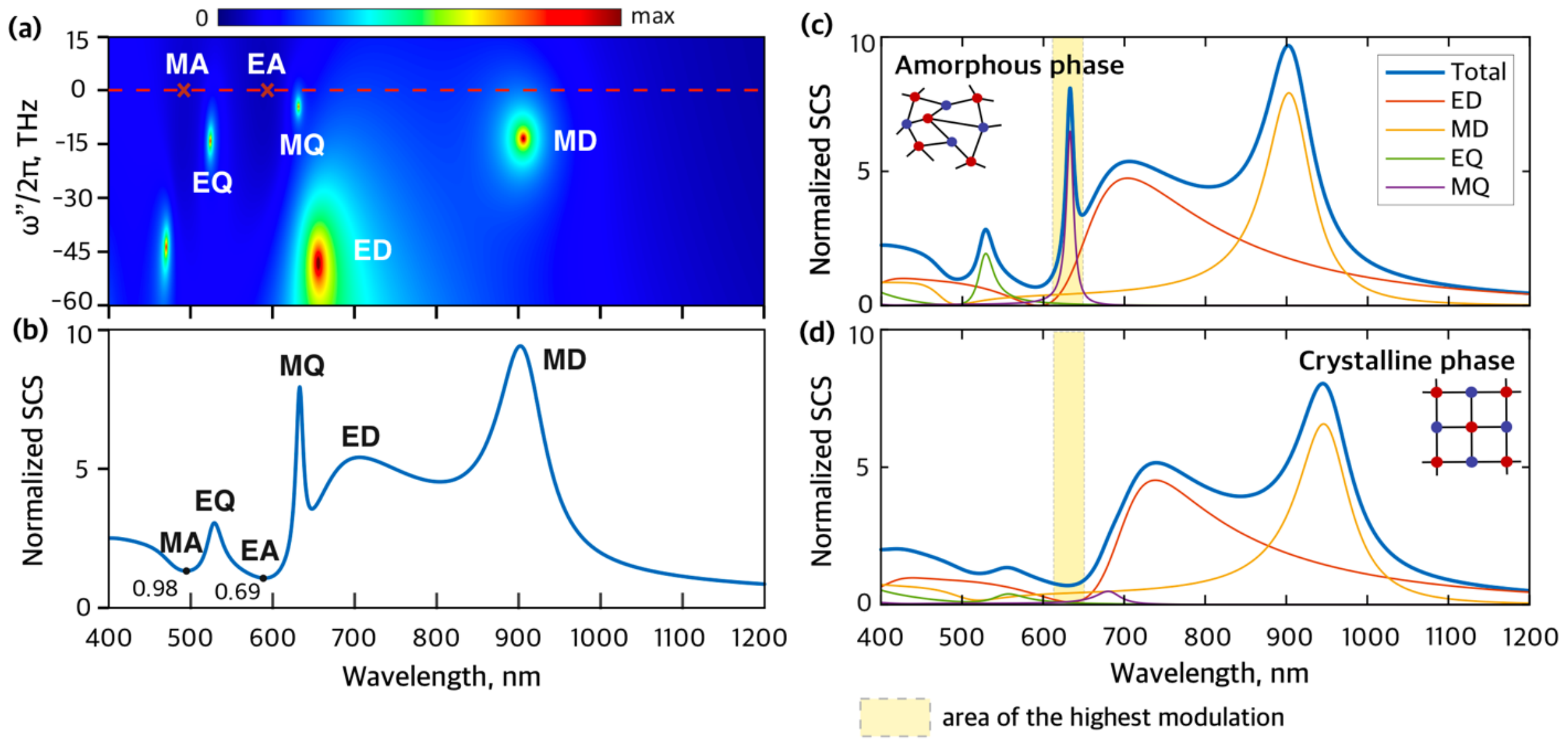

**Figure 2. Passive core-shell nanostructure.** (a) Normalized scattering cross-section (SCS) of a passive core-shell NP based on $Sb_2S_3$ shell and semiconductor core depending on wavelength ($\lambda = 2\pi c/\omega'$)

and imaginary part of frequency. The bright points (MD, ED, MQ, EQ) on the map represent poles of SCS, i.e., eigenfrequencies of the NP. The red marks (EA, MA) correspond to anapoles of the NP. (b) Normalized SCS at zero imaginary frequency depending on wavelength. (c,d) Multipole decomposition of normalized SCS of the core-shell NP in the amorphous (c) and crystalline (d) phase. The yellow area emphasizes the spectral range of the highest modulation of SCS due to the phase change.

The map of SCS of a passive ($g$ = 0) core-shell nanostructure in an amorphous phase depending on the imaginary frequency $\omega''/2\pi$ and wavelength $\lambda = 2\pi c/\omega'$ is shown in Fig. 2(a). Due to passivity, the poles associated with the electric and magnetic multipoles of different order are observed in the lower complex frequency plane. The first four modes are magnetic dipole (MD) with an extracted eigenfrequency $\omega_{\mathrm{MD}}/2\pi$ = 330.6 − 13.25$i$ THz, electric dipole (ED) with $\omega_{\mathrm{ED}}/2\pi$ = 456 − 47.75$i$ THz, magnetic quadrupole (MQ) with $\omega_{\mathrm{MQ}}/2\pi$ = 474.0 – 4.5$i$ THz, and electric quadrupole (EQ) with $\omega_{\mathrm{EQ}}/2\pi$ = 570 – 13.9 THz. The laser threshold can be estimated via the quality factor (Q-factor) of the mode of interest, or in other words, by the amount of gain required to deliver the pole to the real axis[34,38]. Our calculations reveal the MQ mode to have the highest Q-factor of $Q$ = 53 among the first four, which makes this mode a promising candidate for the nanolaser. The red marks in Fig. 2(a) designate the minima of the SCS located at the real axis ( $\omega'' = 0$ ). In literature, these minima are called non-scattering anapoles responsible for optical invisibility or cloaking[30,31,39]. Notably, the electric anapole (EA) is located near the MQ resonance, which facilitates the tunability between MQ and EA states. The poles in the complex frequency plane project into resonances at the real axis, Fig. 2(b). Also, note that anapole states are not eigenmodes of the system and, as such, they cannot lase, unlike poles[31]. We identify the multipole content of the investigated modes and anapoles of the core-shell in the amorphous and crystalline phase by the multipole expansion of the scattering spectra, Fig. 2(c,d). The transformation of the $Sb_2S_3$ shell from amorphous to crystalline state leads to a switch of the scattering response around 633 nm wavelength (yellow area) from MQ resonance to non-scattering EA state.

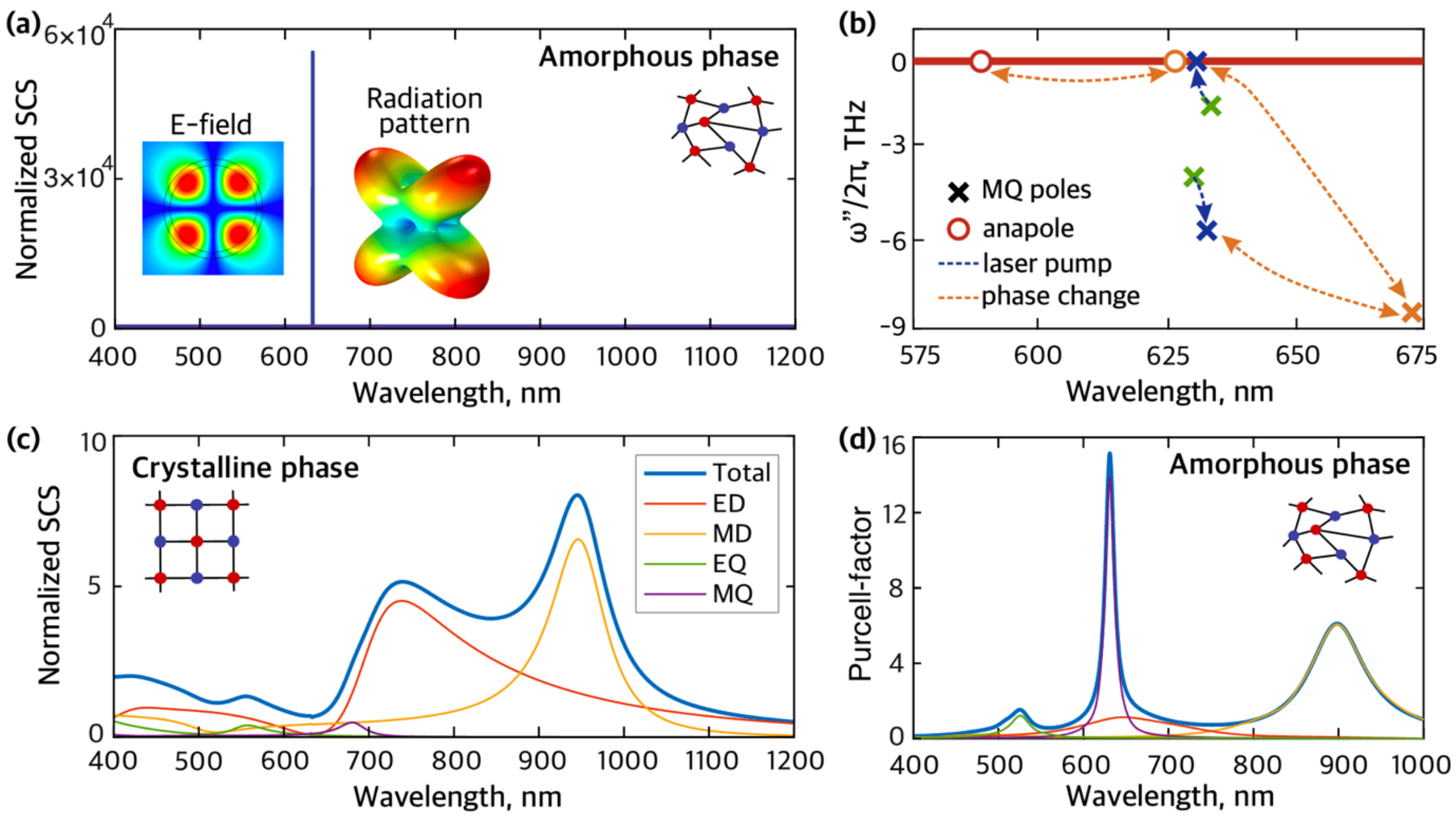

**Figure 3. Active core-shell nanostructure.** Spectral dependence of normalized SCS of the NP with material gain $g$ = 9.5×10$^3$ cm$^{-1}$ in (a) amorphous and (b) crystalline phase of the $Sb_2S_3$ shell and its multipole decomposition. The ultra-high peak of SCS at 633 nm in the amorphous phase corresponds to the laser regime. In the crystalline phase, the pronounced dip is observed at the region around 633 nm. Insert shows the electric field profile and radiation pattern of the lasing mode (MQ). (c) Scheme of dynamics of poles and anapole (scattering zero) of the NP upon the laser pump and phase change. Different colors label poles and anapole corresponding to different states. Green corresponds to the system with the core in the amorphous state and $g$ = 0. Blue denotes the transition of the NP to the nanolaser (the phase remains amorphous) with $g$ = 9.5×10$^3$ cm$^{-1}$ as the laser pump rises. Orange indicates reversible tuning between amorphous and crystalline phases. (d) Purcell factor and its partial contributions from different modes of the NP with amorphous shell depending on wavelength.

Next, we study the transition to lasing in the core-shell nanostructure achieved by introducing the positive material gain. The nanostructure is designed to maximize the overlap between the lasing MQ mode and material gain. The electric field distribution and the radiation pattern are shown in the insert of Fig. 3(a). After introducing a non-zero value of the material gain, the NP modes start interacting with the material dipole transition, and the initial MQ mode splits into two polaritonic modes. These polaritonic modes appear as new distinct poles in the complex frequency plane. Upon increasing the material gain, the poles of the nanostructure shift from their initial positions marked by the green crosses in Fig. 3(c). One of them reaches the real axis when the material gain reaches as high as $g_{th}$ = 9.5×10$^3$ cm$^{-1}$, marking the transition to lasing. The transition is accompanied by a sharpening of the SCS at the wavelength of laser radiation, as shown in Fig. 3(a). Remarkably, pumping of the gain medium does not significantly change the scattering properties of the nanostructure in the crystalline phase that remains cloaked (Fig. 3(b)).

The dynamics of the nanolaser is complicated since the polarization decay rate Γ of the dipole transition in the core is lower than the photonic decay rate of MQ mode $\kappa = 5.65\times10^{13}$ s$^{-1}$. As a result, the dipole moment of the nanolaser gain medium cannot be adiabatically eliminated, which is sometimes referred to as a "superradiant regime"[40,41]. For simulation of the input-output characteristic, we employ an approach that incorporates collective polarization correlations of emitters in the core into the densities of states and frequency-dependent population functions, thereby obtaining the spontaneous and stimulated emission rates in the convenient form of Fermi's "golden rule" integrals[42]. Thus, the spontaneous ($R_{\text{sp}}$) and stimulated ($R_{\text{stim}}$) emission rates into the laser mode are evaluated as:

$$R_{\text{sp}} = \frac{1}{4} G\Gamma \frac{n_2(1-n_1)}{n_2 - n_1} \frac{\kappa + \Gamma - G/2}{(\omega_0 - \omega_{\text{MQ}})^2 + \left(\frac{\kappa + \Gamma - G/2}{2}\right)^2} \tag{2}$$

$$R_{\text{stim}} = \kappa N_p - R_{\text{sp}} \tag{3}$$

where $n_2$ and $n_1$ are occupation numbers of the excited and ground states of dipolar transition, $G = \kappa \frac{g}{g_{\text{th}}} \left[ 1 + 4\left(\frac{\omega_0 - \omega_{\text{MQ}}}{\kappa + \Gamma}\right)^2 \right]$ and the number of photons in the cavity is:

$$N_{\text{p}} = \frac{\Gamma}{\Gamma + \kappa} \frac{n_2(1-n_1)}{n_2 - n_1} \frac{g}{g_{\text{th}} - g} \tag{4}$$

Eqs. (2)–(4) imply that, in the steady-state, the pole cannot cross or even reach the real axis, since, as $g$ approaches $g_{\text{th}}$, the number of photons in the cavity grows infinitely. Also, from Eq. (2), it is evident that the beta-factor, i.e., the ratio of spontaneous emission into the laser mode to the total spontaneous emission into all modes, explicitly depends on the material gain, and therefore, it is not a parameter of the nanolaser.

To evaluate spontaneous emission into non-lasing modes, we have numerically determined the spectrum of Purcell enhancement $F(\omega)$ for a monochromatic emitter in the core and performed its multipole decomposition. The result, shown in Fig. 3(d), is averaged over dipole orientations and positions within the core[43,44]. From the Purcell enhancement spectrum, we were able to rigorously compute the beta-factor at transparency ($g$ = 0) using: $\beta_0 = \frac{\int_0^\infty F_{\text{MQ}}(\omega)\varepsilon_c''(\omega)d\omega}{\int_0^\infty F(\omega)\varepsilon_c''(\omega)d\omega}$, where $F_{\text{MQ}}(\omega)$ is the contribution of the magnetic quadrupole mode to $F(\omega)$ while $\varepsilon_c''(\omega)$ is the imaginary part of the dielectric function $\varepsilon_c''(\omega)$, given by Eq. (1). After calculations, we obtain $\beta_0 = 0.841$. The emission rate into the nonlasing MD, ED and EQ modes is: $R_{\text{sp}}^{\text{nl}} = \frac{1-\beta_0}{4\beta_0} G\Gamma \frac{n_2(1-n_1)}{n_2 - n_1} \frac{\kappa + \Gamma}{(\omega_0 - \omega_{\text{MQ}})^2 + (\kappa + \Gamma)^2/4}$.

Finally, we include the nonradiative decay of excited emitters with the total rate $R_{nr} = Nn_2(1-n_1)/\tau_{nr}$, where $N$ = 1300 is the number of emitters and $\tau_{nr}$ = 1 ns is the nonradiative lifetime.

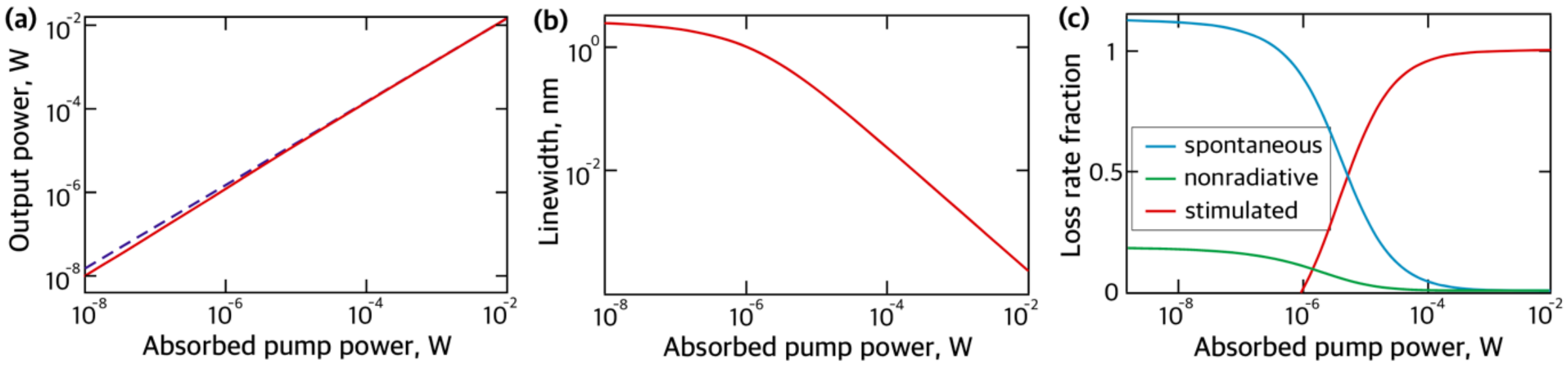

**Figure 4**. **The nanolaser characteristics.** (a) Light input-light output curve of the nanolaser. The blue dashed line corresponds to the characteristic of an ideal thresholdless nanolaser (for comparison). (b) Calculated linewidth of the nanolaser as a function of the pump power. (c) Fractions of spontaneous, stimulated emission rates and nonradiative decay in the total pump rate.

The input-output characteristic of the nanolaser is shown in Fig. 4(a). The log-log plot does not exhibit a distinct kink which makes it similar to thresholdless lasers[45,46]. This can be attributed to the high beta-factor close to 1, and strong Purcell enhancement in the strongly confined Mie resonance mode. The latter makes nonradiative recombination rates relatively small compared to the rates of radiative transitions. In a thresholdless nanolaser, it is impossible to recognize the transition to lasing based only on the input-output curve[47]. To clarify the issue, we have determined the emission linewidth as a function of the absorbed pump power, depicted in Fig. 4(b), which clearly shows the onset of coherence at approximately 1 μW of absorbed pump power. On the plateau below 1 μW, the linewidth of about 2 nm is determined by the polarization dephasing rate of the gain medium, rather than the lifetime of the cavity photons. Upon increase of the absorbed pump power above 1 μW, the laser linewidth decreases inversely proportional to the output power, which agrees with the Schawlow-Townes law and experimentally observed linewidth dependence in thresholdless nanolasers[45,46]. As indicated by Fig. 4(c), the line narrowing coincides with the onset of stimulated emission in our structure. At the same time, linewidth narrowing does not guarantee that the statistics of emitted photons would be Poissonian, as it should be for the coherent state[48]. The number of photons in the cavity[47] is estimated as $N_p \approx \sqrt{Ng_{th}/(dg/dn_2)} \approx 23$, which corresponds to the absorbed pump power of 0.2 mW. At such pump power, only 1 out of 54 photons generated in the laser mode originates from spontaneous emission.

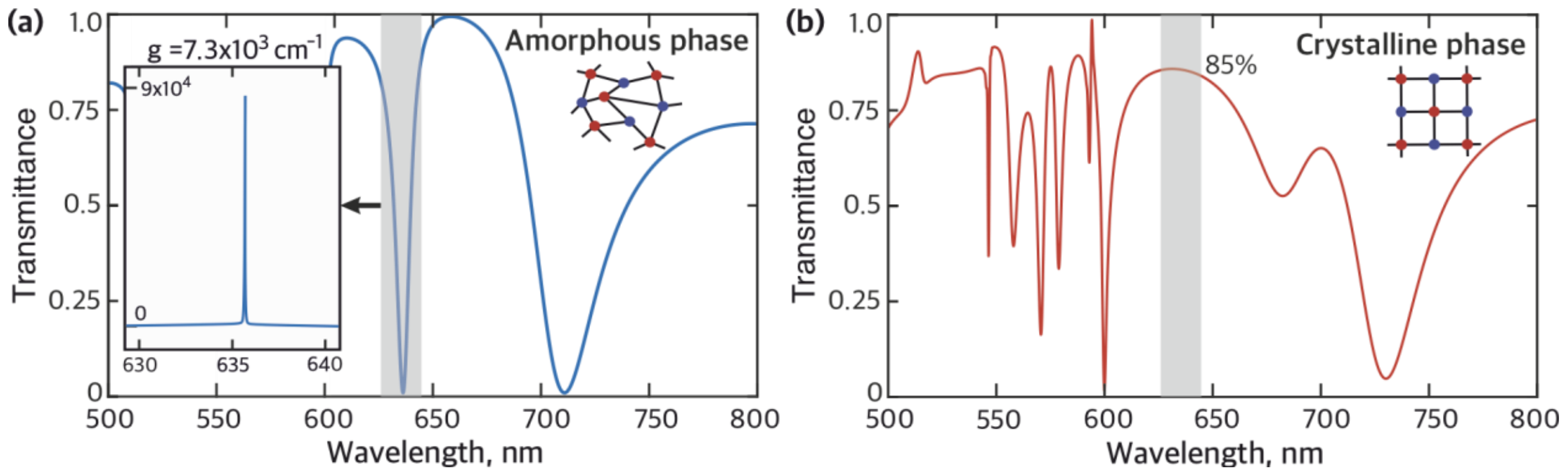

**Figure 5. Cloaking a metasurface.** (a) Transmittance spectrum of a metasurface composed of a square array of core-shell NPs in amorphous phase without gain. The insert shows the transmittance spectrum in the amorphous phase with gain of $7.3\times10^3$ cm$^{-1}$ in the shaded spectral region. (b) Transmittance spectrum of the metasurface in the crystalline phase with the same amount of gain.

Finally, we arrange the core-shell NPs in a square array to study the abovementioned effects in a more realistic case of a metasurface-based transmitter. The period of the metasurface is 600 nm. The NP core and shell radii are 109 nm and 125 nm, respectively. We perform numerical calculations of the transmittance spectra of the metasurface in COMSOL Multiphysics with and without gain in amorphous and crystalline phases, Fig. 5. The passive metasurface transmittance spectrum in the amorphous phase possesses a pronounced resonance in the vicinity of MQ resonance (see Fig. 5(a), shaded area). The resonance position is slightly shifted to the longer wavelengths due to interparticle interactions. The Q-factor of the resonance is increased with respect to the single NP MQ resonance, which is explained by the collective coupling of the NPs. The increased Q-factor enables a lower gain lasing threshold. The insert in Fig. 5(a) shows the transmittance spectrum of the amorphous phase metasurface with gain of $7.3\times10^3$ cm$^{-1}$, which is around 25% less than the initial gain of the single NP. The transmittance reveals an ultrasharp peak related to the nanolaser regime when the pole of the metasurface scattering matrix reaches the axis of real frequencies. However, after the phase change from amorphous to crystalline, the peak turns to a broad non-resonant response characterized by approximately 85% transmittance from 625 nm to 645 nm wavelength, Fig. 5(b). This allows us to conclude that the metasurface based on core-shell NPs can be reversibly switched from the emitting nanolaser regime to the transparent cloaking state.

## Conclusions

In this work, we have proposed a design of a nanolaser based on a semiconductor nanoparticle with gain coated with a film of a phase change material switchable between lasing and cloaking modes at the same operating frequency. The cloaking phase of the nanostructure is associated with the anapole state. The operation characteristics of the nanolaser are rigorously investigated. The nanolaser exhibits characteristics of thresholdless lasers without a pronounced kink in the input-light output curve due to the high betta factor and Purcell factor. Notably, owing to the strong cavity losses, the laser operates in the supperradiant regime, in which the macroscopic dipole moment of the gain medium allows to reach the threshold at lower gain. The distinct feature of this regime is narrower linewidth, which is determined by the emitters dephasing rate rather than the cavity quality factor. We have also designed a reconfigurable metasurface composed of lasing-cloaking metaatoms with the ability to switch from laser radiation to a nonscattering state in a reversible manner. These results can be useful for various photonic and nano-optical systems, especially in cases where the light source has to be able to switch to a transparent state.

# Cloaking a nanolaser: Supplementary information

Sergey Lepeshov[1], Andrey Vyshnevyy[2], and Alex Krasnok[3,*]

[1]*Department of Physics and Engineering, ITMO University, St. Petersburg, Russia*

[2]*Center for Photonics and 2D materials, Moscow Institute of Physics and Technology, Dolgoprudny, 141700, Russia*

[3]*Department of Electrical and Computer Engineering, Florida International University, Miami, FL 33174, USA*

$Sb_2S_3$ is a prospective phase-change material with a high dielectric constant, low loss, and strong tunability in the visible range[32,33]. The permittivity dispersion of $Sb_2S_3$ is shown in Figure S1, and it indicates the broadband transparency of this material above the wavelength of 600 nm in the amorphous phase.

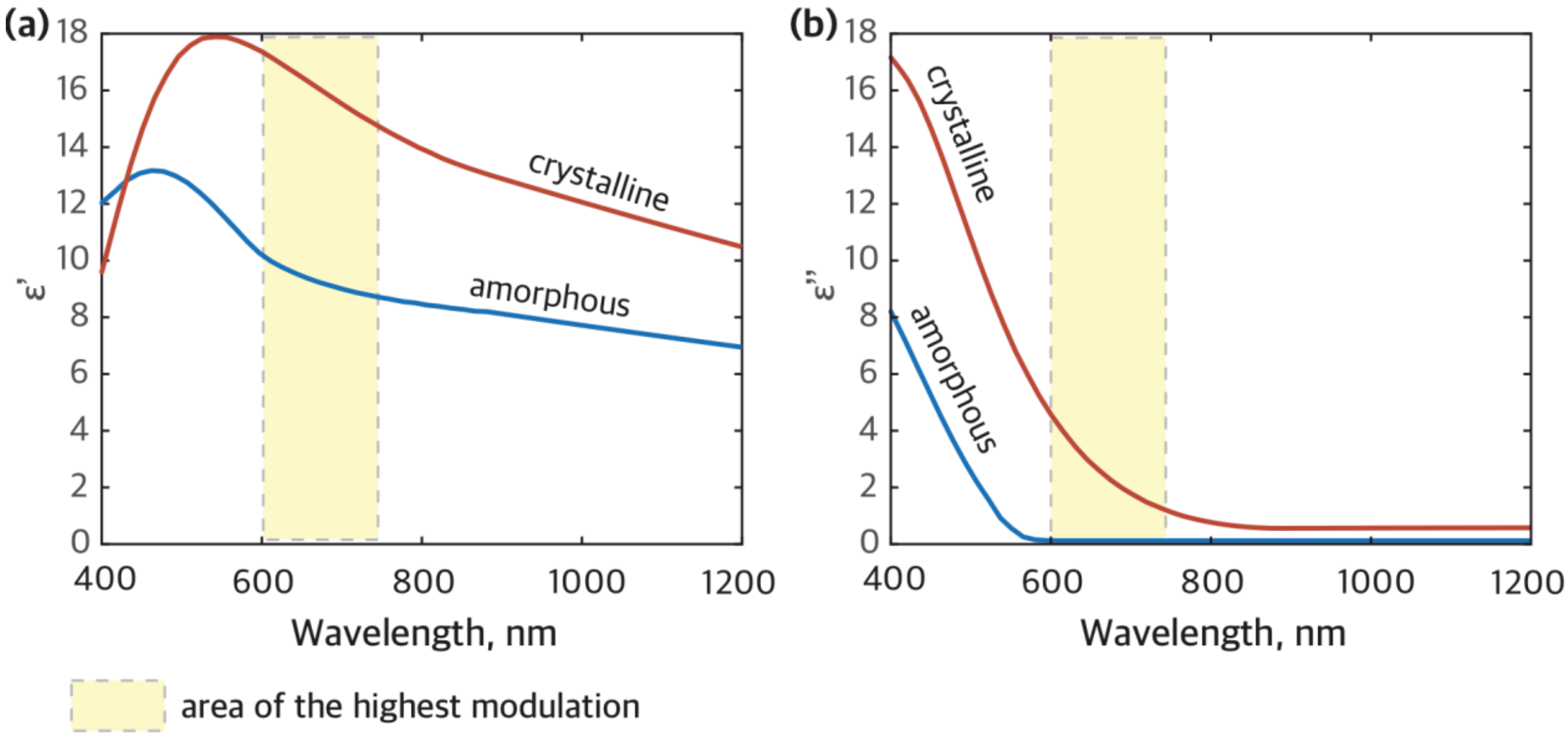

**Figure S1. – (a)** Real and **(b)** imaginary dielectric permittivity of $Sb_2S_3$ in amorphous (blue) and crystalline (red) phases.